\documentclass[aps,superscriptaddress,,twocolumn,showpacs]{revtex4-1}
\pdfoutput=1
\usepackage{color}
\usepackage{amsmath, amsthm, amsfonts}    
\usepackage{amssymb}
\usepackage{mathptmx} 
\usepackage{dsfont}
\usepackage{cancel}
\usepackage{tikz}
\usepackage[unicode=true,pdfusetitle,
 bookmarks=true,bookmarksnumbered=false,bookmarksopen=false,
 breaklinks=true,pdfborder={0 0 0},backref=false,colorlinks=true,citecolor=blue]{hyperref}
\usepackage{graphicx}   
\usepackage[caption=false]{subfig}       
\usepackage{bm}            
\usepackage[normalem]{ulem}  

\newlength\imagewidth
\newlength\imagescale
\def\be{\begin{eqnarray}}
\def\ee{\end{eqnarray}}
\def\r{{\bf r}}

\def\E{{\bf E}}

\def\im{{\rm i}}

\definecolor{JOT-color}{named}{blue}
\definecolor{CSF-color}{named}{orange}

\begin{document}

\title{Multipolar analysis on the 
 helicity conservation in light scattering by single particles}

 \title{Unveiling unusual dipolar spectral regimes of dielectric Mie spheres from the helicity conservation}

\title{Unveiling dipolar regimes  of dielectric Mie spheres beyond the quadrupole resonances from helicity conservation}

\title{Unveiling dipolar spectral regimes of large dielectric Mie spheres from helicity conservation}

\author{Jorge Olmos-Trigo}
\email{jolmostrigo@gmail.com}
\affiliation{Donostia International Physics Center (DIPC),  20018 Donostia-San Sebasti\'{a}n,  Spain}

\author{Diego R. Abujetas}
\affiliation{Donostia International Physics Center (DIPC),  20018 Donostia-San Sebasti\'{a}n,  Spain}
\affiliation{Instituto de Estructura de la Materia (IEM-CSIC), Consejo Superior de Investigaciones Científicas,  Serrano 121, 28006 Madrid, Spain}

\author{Cristina Sanz-Fern\'andez}
\affiliation{Centro de F\'{i}sica de Materiales (CFM-MPC), Centro Mixto CSIC-UPV/EHU,  20018 Donostia-San Sebasti\'{a}n, Spain}

\author{Nuno de Sousa}
\affiliation{Donostia International Physics Center (DIPC),  20018 Donostia-San Sebasti\'{a}n,  Spain}

\author{Jos\'e A. S\'anchez-Gil}
\affiliation{Instituto de Estructura de la Materia (IEM-CSIC), Consejo Superior de Investigaciones Científicas,  Serrano 121, 28006 Madrid, Spain}

\author{Juan Jos\'e S\'aenz}
\affiliation{Donostia International Physics Center (DIPC),  20018 Donostia-San Sebasti\'{a}n,  Spain}
\affiliation{IKERBASQUE, Basque Foundation for Science, 48013 Bilbao,  Spain}

\begin{abstract}
Controlling the electromagnetic helicity and directionality of the light scattered by dielectric particles is paramount to a variety of phenomenology of interest in all-dielectric optics and photonics.
In this Letter, we show that the conservation of the electromagnetic helicity in the scattering by high-index dielectric Mie spheres can be used as a probe of \emph{pure-multipolar} spectral regions, particularly, of dipolar nature, beyond its presumed spectral interval. This finding reveals that the dipolar behaviour is not necessarily limited to small particles, in striking  contrast to the current state of the art. Interestingly, we demonstrate that the \emph{optimum forward light scattering} condition, predicted for a particular nano-sphere in the limit of small particle, is fulfilled in fact  at a fixed ratio between the incident wavelength and particle's size for an infinite number of refractive indexes. 
\end{abstract}

\maketitle
Gustav Mie  presented in  1908  his most relevant contribution to the electromagnetic (EM) theory  by solving the scattering of a plane wave by a spherical particle~\cite{mie1908beitrage}. At its origins,  the aim of this work  consisted on a fundamental theoretical explanation of the colouration of metals, particularly gold colloids~\cite{mishchenko2008gustav}.

As time passed by, the interest on which is nowadays referred to as \emph{Mie theory} became notorious and was later extended to 
magnetic spheres presenting a non-zero relative permittivity $\epsilon$ and permeability $\mu$~\cite{kerker1983electromagnetic}. In that work, Kerker et al.~predicted a \emph{perfect-zero} optical backscattering condition, regardless of the sphere's size, when $\epsilon = \mu$ is satisfied. A few years later, the previous condition was linked to the restoration of a  non-geometrical symmetry: the  EM duality~\cite{fernandez2013electromagnetic}, first introduced by Calkin in 1965~\cite{calkin1965invariance}. In the latter,  Calkin showed that the conserved quantity related to the EM duality symmetry is the EM helicity ($\Lambda$), defined as the projection of the total angular momentum  onto the linear momentum of the wave.
However, there are no magnetic materials ($\mu \neq 1$) at optical frequencies and, then, the EM duality restoration together with its signatures (EM helicity conservation and absence of backscattered light) could not be experimentally verified.

In striking contrast, high refractive index (HRI) dielectric nano-spheres (with $\mu=1$) present strong magnetic and electric dipolar resonances in the visible~\cite{kuznetsov2012magnetic, kuznetsov2016optically}, as well as in telecom and
near-infrared frequencies~\cite{garcia2011strong}. The interference between these electric and magnetic dipolar modes, or equivalently, the first electric and magnetic  Mie coefficients~\cite{bohren2008absorption}, is entirely embedded in the  asymmetry parameter, $g$~\cite{bohren2008absorption}. When the first  Mie coefficients are identical, at the so-called  first Kerker condition~\cite{nieto2011angle}, $g$ is maximized in the  electric \emph{and magnetic} dipolar regime~\cite{olmos2020optimal}\footnote{An electric \emph{and magnetic} dipolar response will be assumed throughout the text.}, leading to the  zero optical backscattering  condition~\cite{geffrin2012magnetic,person2013demonstration,fu2013directional}. 
For incoming beams with well-defined EM helicity ($\Lambda_{\rm{inc}} = \sigma$, with $\sigma = \pm1$), a direct relation between $\Lambda$ (after scattering) and $g$ arises, namely, $\langle \Lambda \rangle = 2 \sigma g$~\cite{olmos2019asymmetry}, and this absence of backscattered light can be likewise derived from EM helicity conservation~\cite{zambrana2013duality}.

Interestingly, an \emph{optimum forward light scattering} was found at the first Kerker condition for a nano-sphere with given refractive index contrast ${\rm{m}} \approx 2.45$~\cite{luk2015optimum,zhang2015dielectric}.
Such nano-spheres are often referred to as \emph{dual} scatterers~\cite{zambrana2013dual,olmos2019enhanced}, i.e., particles that are invariant under EM duality transformations~\cite{fernandez2013electromagnetic} and, thus, preserve the EM helicity after scattering. Nonetheless, the ideal mapping from the scattering by magnetic spheres satisfying $\epsilon= \mu$ onto the scattering by dielectric spheres can only be achieved in a \emph{multipolar scattering process}~\footnote{We refer to as \emph{multipolar scattering process} whereas several (more than one) multipolar modes are involved in the scattering efficiency.}, when every pair of electric and magnetic Mie coefficients are identical, i.e.,  $a_l = b_l \; \forall \: l$, where $l$ the multipole order. The latter condition has been widely conjectured to be unattainable~\cite{zambrana2013dual,zambrana2013duality, abdelrahman2017broadband}, although an actual demonstration remains  unexplored. In this vein, a  \emph{near-duality} spectral region, apparently satisfying $a_l \approx b_l \; \forall l$, was predicted far beyond the dipolar regime in which a broadband near-zero optical backscattering is achievable~\cite{abdelrahman2017broadband}.

In this Letter, we demonstrate that the ideal mapping from $\epsilon = \mu$ onto the  scattering by dielectric Mie spheres is precluded due to a fundamental  property of the Bessel functions. The latter property is  general since it does not depend on the refractive index contrast, multipole order, incoming polarization or ratio between the incident wavelength and particle's size, and specifically shows that if $a_j = b_j$ then $a_l \neq b_l \; \forall \:  l \neq j$. Consequently, the absence of backscattered light and the EM helicity conservation cannot be \emph{ideally} achieved in a multipolar scattering process. Nevertheless and 
as a result of our proof, we show that the almost entirely conservation of the EM helicity ($ \langle \Lambda \rangle \approx 1$) implies the existence of \emph{pure-multipolar} spectral regions~\footnote{We refer to as \emph{pure-multipolar} scattering processes those that can be described with just one order $l$, e.g. a pure dipolar process.}, specifically of dipolar nature, where the zero optical backscattering condition can be fulfilled for micro-sized spheres.  
Particularly, we determine that the optimum forward light scattering condition arises quasi-periodically for an infinite number of refractive indexes at a fixed ratio between the incident wavelength and particle's size in unexpected dipolar spectral regimes.

In addition, we expose that at the so-called near-duality spectral region~\cite{abdelrahman2017broadband}, in which a broadband nearly-zero optical backscattering is achievable, the EM helicity is not conserved, showing that the absence of backscattered light is not (always) a reliable signature of the EM duality restoration.
Our results clearly  illustrate then that  $\langle \Lambda \rangle \approx  1$ can be used as a straightforward probe of dipolar regimes. Moreover, we unveil that these  arise in unusual spectral regimes beyond the magnetic and electric quadrupole resonances, opening new insights in the study of the scattering of light from dielectric Mie spheres. The latter can be summarized as follows: the concept of \emph{small particle}~\cite{garcia2011strong,kuznetsov2012magnetic, kuznetsov2016optically} is sufficient, but not necessary, in order to assume a dipolar optical response. 

Mie theory~\cite{mie1908beitrage} gives the exact analytical solution of Maxwell's equations for  a spherical particle in an homogeneous medium under plane wave  illumination. It allows to write the scattering efficiency
of the particle as~\cite{bohren2008absorption}
\begin{equation}\label{scattering}
Q_{\rm{sca}} = \frac{2}{x^2} \sum_{l= 1}^{\infty} \left( 2l +1 \right) \left(  |a_l|^2 + |b_l|^2  \right),
\end{equation}
where $Q_{\rm{sca}} =  \sigma_{\rm{sca}} / \pi R^2$, being $\sigma_{\rm{sca}}$ the scattering cross section and $R$ the radius of the particle. Here, $x = kR$ is the size parameter, where $k$ = ${\rm{m}}_{\rm{h}} k_0 = {\rm{m}}_{\rm{h}} \left(2 \pi  / \lambda_0 \right)$, being $\lambda_0$ the incoming wavelength in vacuum and ${\rm{m}}_{\rm{h}}$ is the refractive index of the external medium. The scattering properties depend on the refractive index contrast between the particle and the external medium, defined as ${\rm{m}} = {\rm{m}}_{\rm{p}}/{\rm{m}}_{\rm{h}}$, where ${\rm{m}}_{\rm{p}}$ is the refractive index of the particle. The electric and magnetic Mie coefficients, $a_l$ and $b_l$, respectively,  are given in terms of the scattering phase-shifts~\citep{hulst1957light} by
\begin{align}\label{Mie_tot}
a_l = \im \sin \alpha_l e^{-\im \alpha_{l}} && \text{and} && b_l = \im \sin \beta_l e^{-\im \beta_{l}}, 
\end{align}
where 
\begin{equation}\label{etan}
\tan \alpha_l = -\frac{S'_l({\rm{m}}x)S_l(x) -{\rm{m}} S_l({\rm{m}}x)S'_l(x)}{S'_l({\rm{m}}x)C_l(x) -{\rm{m}} S_l({\rm{m}}x)C'_l(x)},
\end{equation}
and
\begin{equation}\label{mtan}
\tan \beta_l = -\frac{{\rm{m}}S'_l({\rm{m}}x)S_l(x) -S_l({\rm{m}}x)S'_l(x)}{{\rm{m}}S'_l({\rm{m}}x)C_l(x)- S_l({\rm{m}}x)C'_l(x)}.
\end{equation}
Here $S_l(z) = zj_l(z) = \sqrt{\frac{\pi z }{2}} J_{l+\frac{1}{2}}(z)$ and $C_l(z) = zn_l(z)  =  \sqrt{\frac{\pi z }{2}} N_{l+\frac{1}{2}}(z)$ denote the Riccati-Bessel functions, 
where $j_l(z)$ and $n_l(z)$ denote the spherical Bessel and Neuman functions while  $J_{l+\frac{1}{2}}(z)$  and $N_{l+\frac{1}{2}}(z)$ are  the Bessel and Neumann functions, respectively. 

In this scattering phase-shifts notation, the condition $\alpha_l({\rm{m}},x) = \beta_l({\rm{m}},x) \; \forall \: l \Longleftrightarrow a_l({\rm{m}},x) = b_l({\rm{m}},x) \; \forall \: l$ would imply the restoration of the EM duality and the conservation of its associated signature, the EM helicity, which would lead to the perfect-zero optical backscattering condition in an arbitrary multipolar scattering process~\cite{zambrana2013duality}. For each multipole order $l$ and according to Eqs.~\eqref{etan} and \eqref{mtan}, this non-magnetic generalized duality condition requires either $S_l ({\rm{m}}x) = 0$, or $S'_l ({\rm{m}}x) = 0$~\cite{hulst1957light}. Generally, these solutions are embedded in:
\begin{align}\label{Master}
a\ J_{l+\frac{1}{2}}({\rm{m}}x) + b\ {\rm{m}} x\ J'_{l+\frac{1}{2}}({\rm{m}}x) = 0,
\end{align}
where $a,b \in \mathbb{R}$.
Notice that when $a = 1$ and $b = 0$ the node of the first kind emerges, i.e., $S_l ({\rm{m}}x) = 0$, while for $a = 1$ and $b =2$ the node of the second kind arises, i.e., $S'_l ({\rm{m}}x) = 0$.
At this point, let us make use of the following Lemma~\citep{watson1995treatise}:
\begin{enumerate}
\item  \label{Baricz}  When $v > -1$ and $a,b \in \mathbb{R}$ such as $a^2 + b^2 \neq 0$, then no function of the type  $aJ_{v}(z) + bz J'_{v}(z) = 0$ can have a repeated zero other than $z = 0$.
\end{enumerate} 
Since the trivial solution ${\rm{m}}x = 0$ implies no particle, this Lemma necessarily means that Eq.~\eqref{Master} cannot be formally satisfied $\forall  \: l$ for a fixed size parameter ${\rm{m}}x$. On physical grounds, this phenomenon implies that, when $a_j({\rm{m}},x) = b_j({\rm{m}},x)$, no other pair of electric and magnetic Mie coefficients can be identical, $a_l({\rm{m}},x) \neq b_l({\rm{m}},x) \;  \forall  \: l \neq j$. Lemma~\ref{Baricz} unveils a property of a dielectric {Mie sphere} that has been broadly conjectured~\cite{zambrana2013dual, zambrana2013duality, abdelrahman2017broadband} but not yet demonstrated: the EM helicity cannot be totally conserved in a multipolar scattering process. This fact precludes the perfect-zero optical backscattering condition in a cylindrically symmetric target~\cite{zambrana2013duality} and the \emph{ideal} restoration of the  EM duality in a multipolar scattering process~\cite{fernandez2013electromagnetic}, regardless of the incoming polarization, multipole order, size parameter $x$, and refractive index contrast. Thus, we have revealed an inherent  property of the so-called Mie theory~\cite{mie1908beitrage}.

\begin{figure}[t!]
\includegraphics[width=1 \columnwidth]{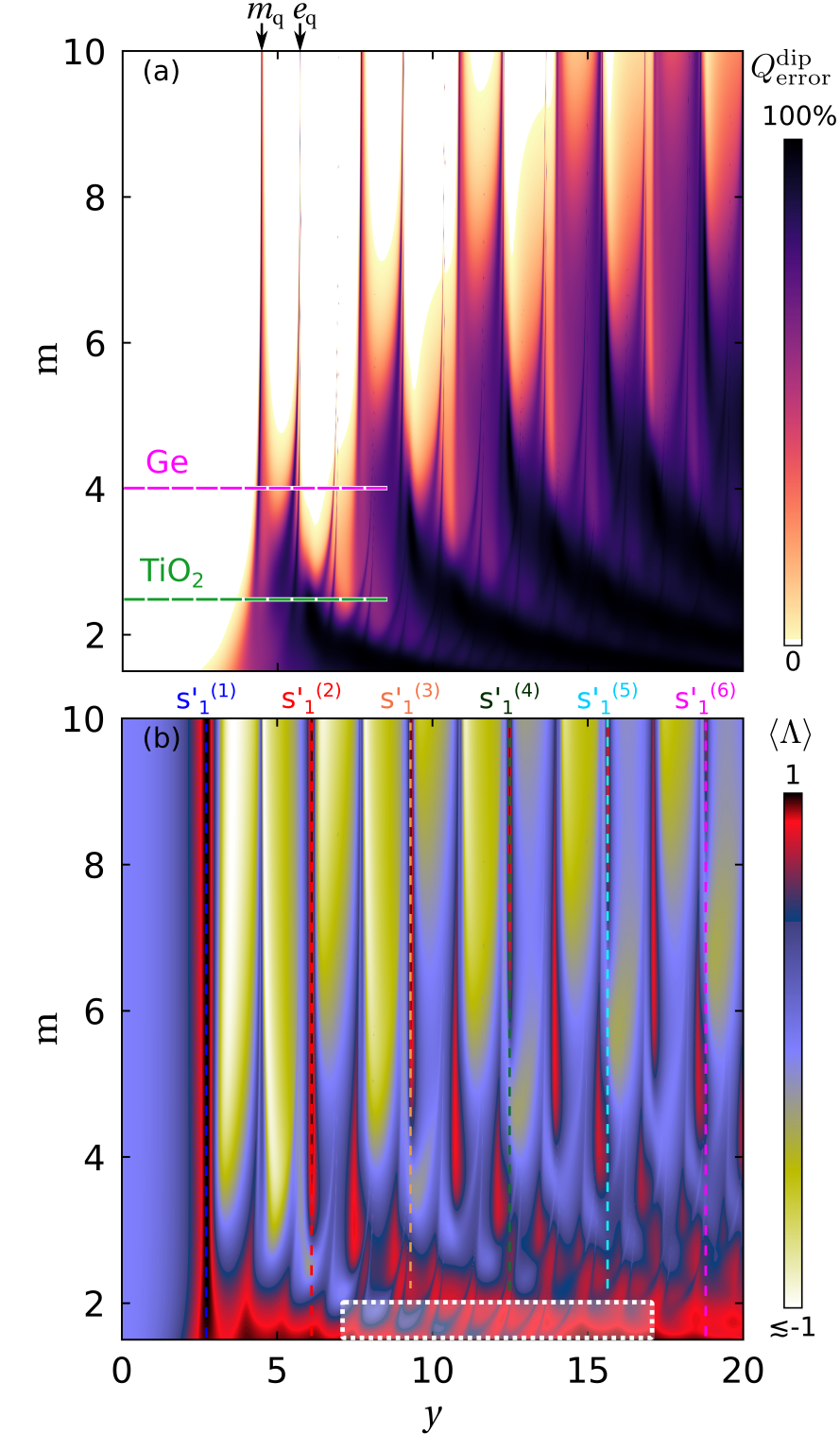}
\captionsetup{justification= raggedright}
\caption{(a) Percentage error of assuming a dipolar response, i.e., $Q^{\rm{dip}}_{\rm{error}} = |Q^{\rm{dip}}_{\rm{sca}} /Q_{\rm{sca}} -1 |\%$ vs the $y={\rm{m}}x$ size parameter and refractive index contrast  ${\rm{m}}$. The dipolar spectral regime corresponds to white region. (b) Color map of the expected value of the EM helicity after scattering by  a dielectric Mie sphere, $\langle \Lambda \rangle$,  under well-defined EM helicity ($\sigma = +1$) plane wave illumination. Notice that  $\langle \Lambda \rangle \approx 1$ at the  $s'^{(q)}_1$ trajectories when the scattering can be described by electric and magnetic dipolar modes. The white dashed transparent rectangle  illustrates the so-called near-duality spectral region~\cite{abdelrahman2017broadband}.}
\label{colormap}
\end{figure}

Nevertheless, the EM helicity can be almost entirely conserved in pure-multipolar spectral regions with well-defined square of the total angular momentum, ${\bf{J}}^2$. Specifically, in  dipolar spectral regimes, as we will shortly see.
Let us explicitly calculate the expected value of the scattered EM helicity arising from a dielectric {Mie sphere}. 
Within a helicity and  angular momentum framework, the incoming beam electric field, $\E_{\text{inc}}$, can be expanded in vector spherical wavefunctions (VSWFs), $\boldsymbol{\Psi}_{lm}^{\sigma}$,  with well-defined helicity,  $\sigma = \pm 1$~\cite{olmos2019enhanced,olmos2019sectoral}. In this basis, 
\be
\E_{\text{inc}} &=& \sum_{\sigma= \pm 1} \E_{\text{inc}}^\sigma, \quad  \E_{\text{inc}}^\sigma = E_0 \sum_{l=1}^{\infty} \sum_{m=-l}^{+l}  C_{lm}^{ \sigma} \boldsymbol{\Psi}_{lm}^{\sigma},
\label{multipolar} 
\ee
where $\boldsymbol{\Psi}_{lm}^{\sigma}$ is defined as
\be
\boldsymbol{\Psi}_{lm}^{\sigma} &=& \frac{1}{\sqrt{2}} \left[ {\boldsymbol{N}}_{lm} +  \sigma {\boldsymbol{M}}_{lm}  \right], \label{V1} \\
{\boldsymbol{M}}_{lm} &\equiv & j_l(kr)\boldsymbol{X}_{lm},  \quad
{\boldsymbol{N}}_{lm} \equiv \frac{1}{k} \boldsymbol{\nabla} \times {\boldsymbol{M}}_{lm}, \\
\boldsymbol{X}_{lm} &\equiv& \frac{1}{\sqrt{l(l+1)}} {\bf{L}} Y_l^m (\theta,\varphi). \label{V3}
\ee
Here, $\boldsymbol{M}_{lm}$ and $\boldsymbol{N}_{lm}$  are Hansen's multipoles, $\boldsymbol{X}_{lm}$ denotes the vector spherical harmonic~\cite{jackson1999electrodynamics}, $ j_l(kr)$ are the spherical Bessel functions (well-defined at $r = 0$),  $Y_l^m$ are the spherical harmonics, $C_{lm}^{ \sigma}$ are the incident coefficients characterizing the nature of the incoming wave, and $ {\bf{L}} =  \left\{ -\im \r \times \boldsymbol{\nabla}\right\} $ is the orbital angular momentum operator.
Let us recall that the multipoles $ \boldsymbol{\Psi}_{lm}^{\sigma} $ 
can be built following the standard rules of angular momentum addition~\cite{edmonds2016angular}
as simultaneous eigenvectors of $\bf{J}^2$ and $J_z$, with $\bf{J} = {\bf{L}} + {\bf{S}}$, being ${\bf{S}} = - \im \times$ the  spin angular momentum  operator. The multipoles $ \boldsymbol{\Psi}_{lm}^{\sigma} $ are then simultaneous eigenvectors of $\bf{J}^2$, $J_z$~\cite{edmonds2016angular} and the helicity operator $\boldsymbol{\Lambda} = (1/k) \boldsymbol{\nabla} \times$~\cite{calkin1965invariance}, with eigenvalues $l(l+1)$, $m$ and $\sigma$, respectively.

\begin{figure}[t!]
\includegraphics[width=1 \columnwidth]{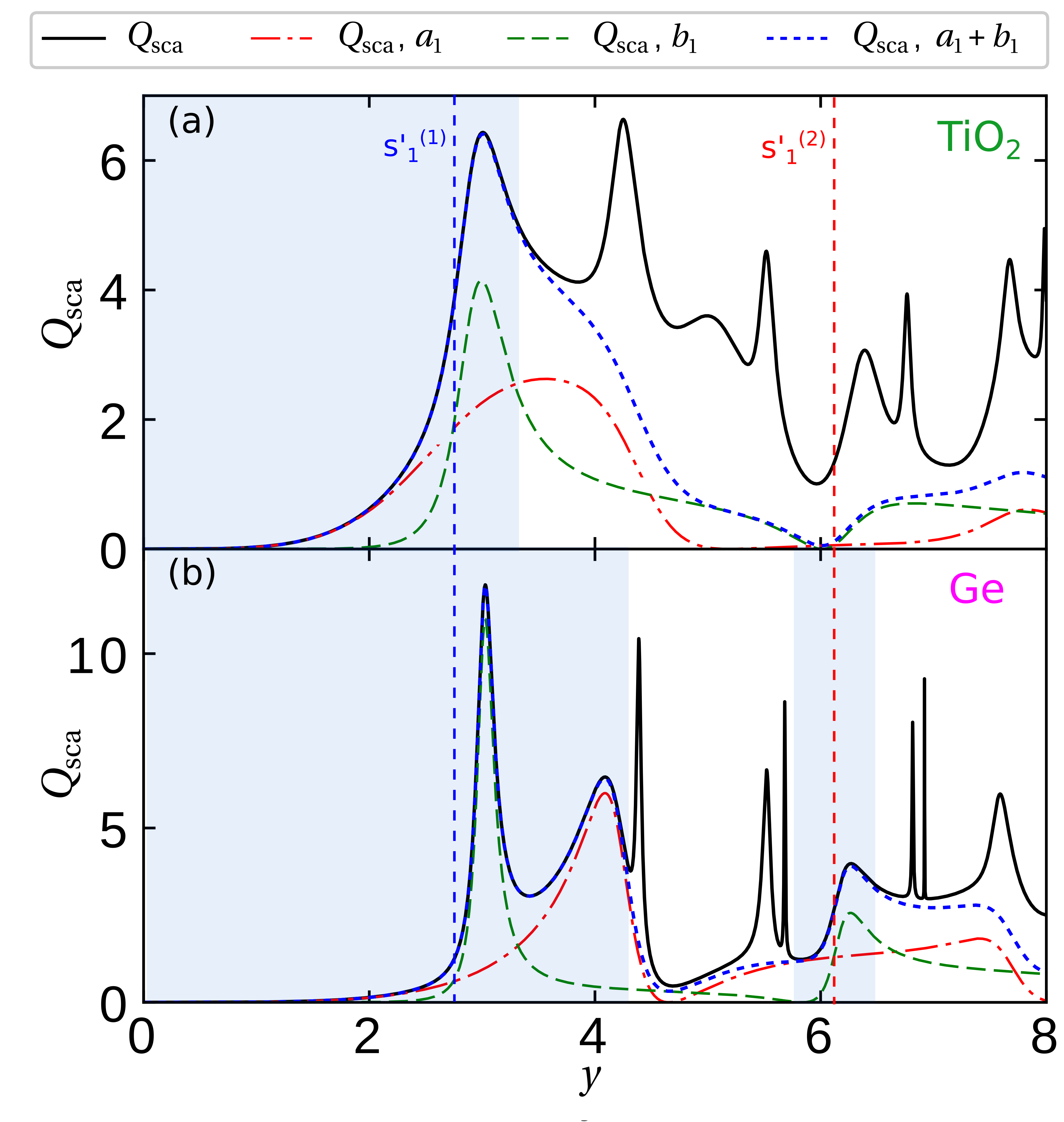}
\captionsetup{justification= raggedright}
\caption{Scattering efficiency from: (a) TiO$_2$-like sphere with ${\rm{m}} = 2.5$ and (b) Ge-like sphere  with ${\rm{m}} = 4$. Blue regions illustrate the spectral regimes that are essentially described by a dipolar optical response. The first and second multipolar Kerker conditions are willfully depicted in order to show that just $s'^{(1)}_1$ and $s'^{(2)}_1$  lead to EM helicity preserving for HRI spheres (${\rm{m}}>3.5$), according to Fig.~\ref{colormap}b.}
\label{Pure}
\end{figure}

When a dielectric {Mie sphere} is  centered at the origin ($r = 0$), the AM number $m$ is preserved in the scattering process due to the rotational invariant symmetry of the system.  The scattered fields outside the sphere, $\E_{\text{sca}} = \E_{\text{sca}}^+ + \E_{\text{sca}}^-$, can be written in terms of ``outgoing'' VSWFs, $\bm{\Phi}_{lm}^{\sigma'}$ (defined as in Eq.~\eqref{V1} replacing $j_l(kr)$ by the outgoing spherical Bessel functions $h_l(kr)$), as
\be 
\E_{\text{sca}}^\sigma &=& E_0 \sum_{l=1}^{\infty} \sum_{m=-l}^{+l}  D_{lm}^{ \sigma} \boldsymbol{\Phi}_{lm}^{\sigma}, \nonumber \\
\begin{pmatrix} D_{lm}^+ \\ D_{lm}^- \end{pmatrix} &=&  
-\begin{pmatrix} [a_l+b_l] & -[a_l-b_l] \\ [a_l-b_l] & -[a_l +b_l]  \end{pmatrix} 
\begin{pmatrix} C_{lm}^+ \\   C_{lm}^- \end{pmatrix}. \label{helimatrix}
\label{multipolar_scatter} 
\ee
In this framework, it can be seen that the expected value of the EM helicity of the scattered field is generally given by
\begin{equation} \label{hel_gen}
\langle \Lambda \rangle =
\frac{ \langle \bm{E}_{\rm{sca}}^* \cdot \left(  \bm{\Lambda} \bm{E}_{\rm{sca}} \right) \rangle } 
       {\langle \bm{E}_{\rm{sca}}^* \cdot \bm{E}_{\rm{sca}} \rangle }
=  \frac{\sum_{l=1}^\infty \sum_{m=-l}^{+l} \left[ |D_{lm}^+|^2 - |D_{lm}^-|^2 \right] }{\sum_{l=1}^\infty \sum_{m=-l}^{+l} \left[ |D_{lm}^+|^2 + |D_{lm}^-|^2 \right] }.
\end{equation}

Equation~\eqref{hel_gen} is a general result showing the expected value of the EM helicity after scattering for an arbitrary incident beam. Particularly, when the sphere is illuminated by a circularly polarized plane wave with helicity $\sigma$~\cite{olmos2019enhanced}, or by a cylindrically symmetric beam~\cite{zambrana2013dual} (eigenvector of both $\bm{\Lambda}$ and $J_z$ operators, with eigenvalues $\sigma$ and $m$, respectively), the scattered field is a combination of multipolar modes with fixed $m$. As a result, it can be shown that
\begin{align} \label{hel_mean} 
\langle \Lambda \rangle = \sigma \frac{1-T}{1+T}, &&\text{where} &&  T = \frac{\sum_{l=|m|}^\infty  \left(2l+1 \right)   \left| C_{lm}^+  \right|^2   |a_l -b_l|^2}{\sum_{l=|m|}^\infty \left(2l+1 \right)   \left| C_{lm}^+  \right|^2 |a_l + b_l|^2 }
\end{align}
is the helicity transfer function~\cite{zambrana2013dual}.

When $T$ goes to zero, the particle is said to be dual~\cite{zambrana2013dual} and the scattered EM helicity is identical to the EM helicity of the incoming beam. Nonetheless, as we have previously demonstrated via Lemma~\eqref{Baricz}, the EM helicity cannot be ideally conserved in a multipolar scattering process.
\begin{figure}[t!]
\includegraphics[width=1 \columnwidth]{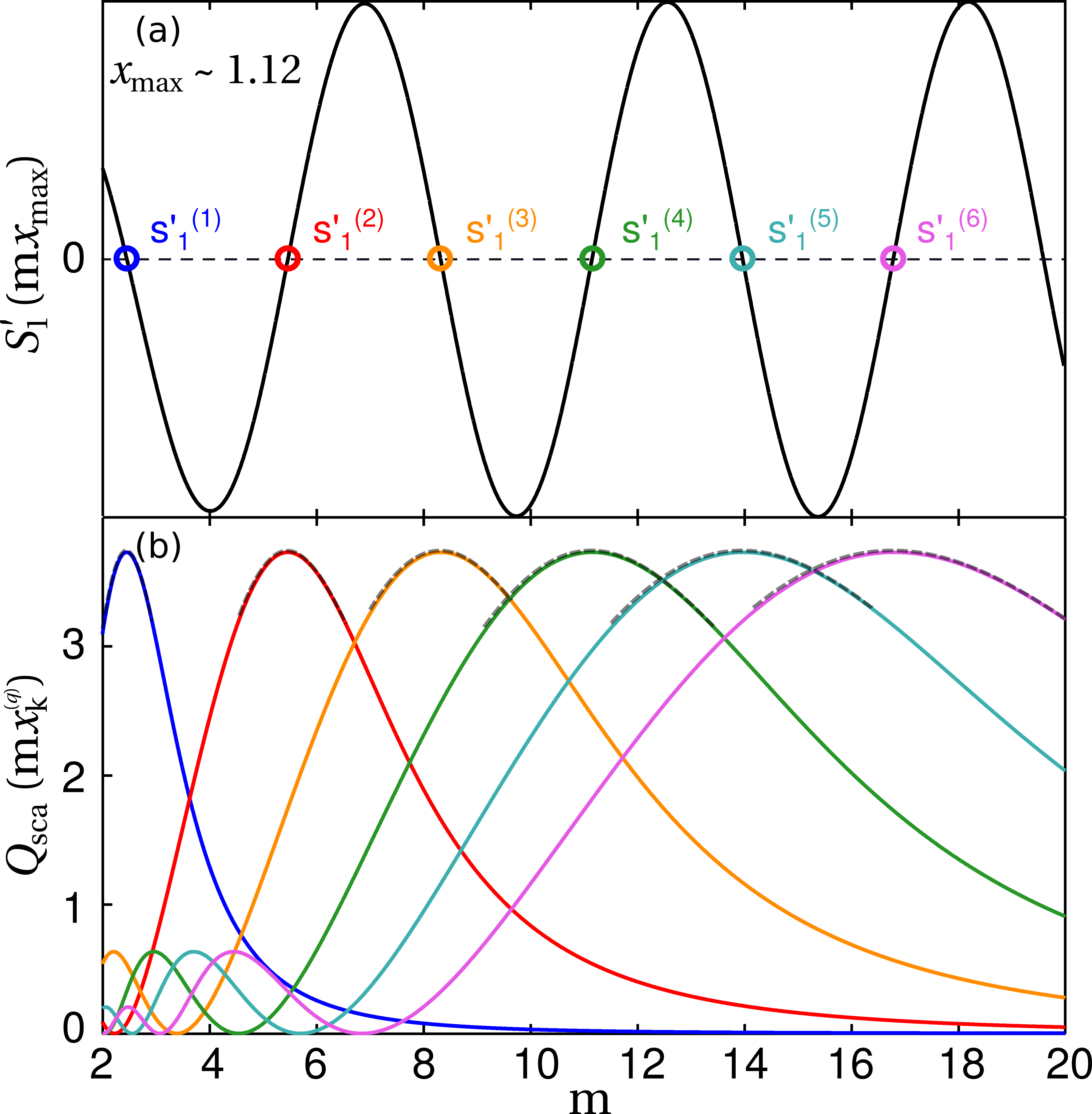}
\captionsetup{justification= raggedright}
\caption{(a) Node of the second kind, $S'_1({\rm{m}}x_{\rm{max}})$,  where $x_{\rm{max}} \approx 1.12$, as a function of the contrast index ${\rm{m}}$. The q-th positive zero of this function, depicted by circles, satisfies the optimum forward light scattering condition. (b) Scattering efficiency, $Q_{\rm{sca}}$, evaluated at the $x^{(q)}_{\rm{k}}$, i.e., the size parameter that leads to  ${s'_{1}}^{(q)}$, as a function of ${\rm{m}}$. The maximum value of $Q_{\rm{sca}}$ is identical to $Q_{\rm{sca}} = 3.75$. Interestingly, at each of these maxima, the full multipolar expansion (gray dashed lines) is essentially dipolar (solid lines).}
\label{lombriz}
\end{figure}
However, the EM helicity can be almost entirely preserved in pure-multipolar spectral regions, particularly, in dipolar regimes, as can be inferred from Fig.~\ref{colormap}. Firstly, we identify dipolar regions (white colors) from Fig.~\ref{colormap}a, where the percentage error of assuming a  dipolar response, i.e., $Q^{\rm{dip}}_{\rm{error}} = |Q^{\rm{dip}}_{\rm{sca}} /Q_{\rm{sca}} -1 |\%$, versus the $y={\rm{m}}x$ size parameter and refractive index contrast  ${\rm{m}}$ is depicted. Notice that $Q^{\rm{dip}}_{\rm{sca}}$ corresponds to Eq.~\eqref{scattering} but retaining only $l=1$ (dipolar contribution). Surprisingly, several dipolar spectral regimes are found far beyond its presumed spectral interval (beyond the magnetic ($m_{\rm{q}}$) and electric quadrupole ($e_{\rm{q}}$) resonances) for HRI spheres with ${\rm{m}} \geq 3.5$. Secondly, we show in Figure~\ref{colormap}b the expected value of the EM helicity after scattering, $\langle \Lambda \rangle$, under well-defined helicity plane wave illumination, $\sigma =+1$ (see Eq.~\eqref{hel_mean}). As can be inferred from the attached color-bar, $\langle \Lambda \rangle \approx 1$ when the $s'^{(q)}_1$ vertical trajectories, corresponding to the $q$-th zeros of $S'_1 ({\rm{m}}x) = 0$~\footnote{{We denote the $q$-th positive zeros of $S_l ({\rm{m}}x) = 0$ and $S'_l ({\rm{m}}x) = 0$ as $s^{(q)}_l$ and $s'^{(q)}_l$, respectively.}}, pass through a dipolar spectral region.
In contrast, $\langle \Lambda \rangle$ is not conserved in a multipolar scattering process as a result of Lemma~\ref{Baricz} since if  $a_l({\rm{m}},x) = b_l({\rm{m}},x)$ then $a_j({\rm{m}},x) \neq b_j({\rm{m}},x) \;  \forall  \: j \neq l$. 
Notice that this phenomenon also applies to the zeros corresponding to the nodes of first kind, $S_1 ({\rm{m}}x) = 0$, according to Lemma~\ref{Baricz}.

Let us now briefly discuss the concept of near-duality~\cite{abdelrahman2017broadband}. Since EM duality implies absence of backscattered light~\cite{zambrana2013duality}, one can be tempted to identify spectral regions with very low backscattering cross sections with near-duality regions, where $a_l \approx b_l \; \forall \:l$.  This condition would imply a nearly-conserved EM helicity since the transfer function of Eq.~\eqref{hel_mean} would be approximately zero. However, as can be inferred from the white semi-transparent rectangle appearing in Fig.~\ref{colormap}b, the expected value of the helicity oscillates between $0.3<\langle\Lambda \rangle<0.9$ in the so-called near-duality spectral region. This contradicts the essence of the EM duality restoration as its signature, the EM helicity, is not preserved. These aforementioned phenomena corroborate that  $\langle \Lambda \rangle \approx 1$ can be used as a straightforward probe of the presence of dual pure-multipolar spectral regions described by electric and magnetic dipolar modes.  
However, it is important to notice that the emergence of dipolar spectral regions does not entail $\langle \Lambda \rangle \approx 1$ in the entire dipolar regime{, which can be observed comparing the white regions of Fig.~\ref{colormap}a with their corresponding regions of Fig.~\ref{colormap}b.}

In order to get a deeper insight into the appearing of these dipolar spectral regimes, we analyze in Fig.~\ref{Pure} the dipolar contribution (blue dashed line) to the total scattering efficiency (black solid line) for two different dielectric-like spheres~\cite{aspnes1983dielectric}: Titanium Oxide (TiO$_2$) with ${\rm{m}} = 2.5$ and Germanium (Ge) with ${\rm{m}} = 4$.  While in the case of TiO$_2$, the dipolar regime (blue background in Fig.~\ref{Pure}a) just emerges in the limit of small particle~\cite{garcia2011strong,kuznetsov2012magnetic,kuznetsov2016optically}, the scattering efficiency arising from the Ge sphere presents an unexpected dipolar spectral regime in the interval given by $5.75< y <6.25$ (narrow blue background in Fig.~\ref{Pure}b), far beyond the magnetic and electric quadrupole resonances.
Let us recall that, at this unusual dipolar spectral regime, the EM helicity is almost entirely preserved at $s'^{(1,2)}_1$ for the Ge sphere while it is not near to be preserved for the TiO$_2$ sphere at $s'^{(2)}_1$ due to the contribution from higher order multipoles, as previously discussed in Fig.~\ref{colormap}b.

\begin{figure}[t!]
\includegraphics[width=1 \columnwidth]{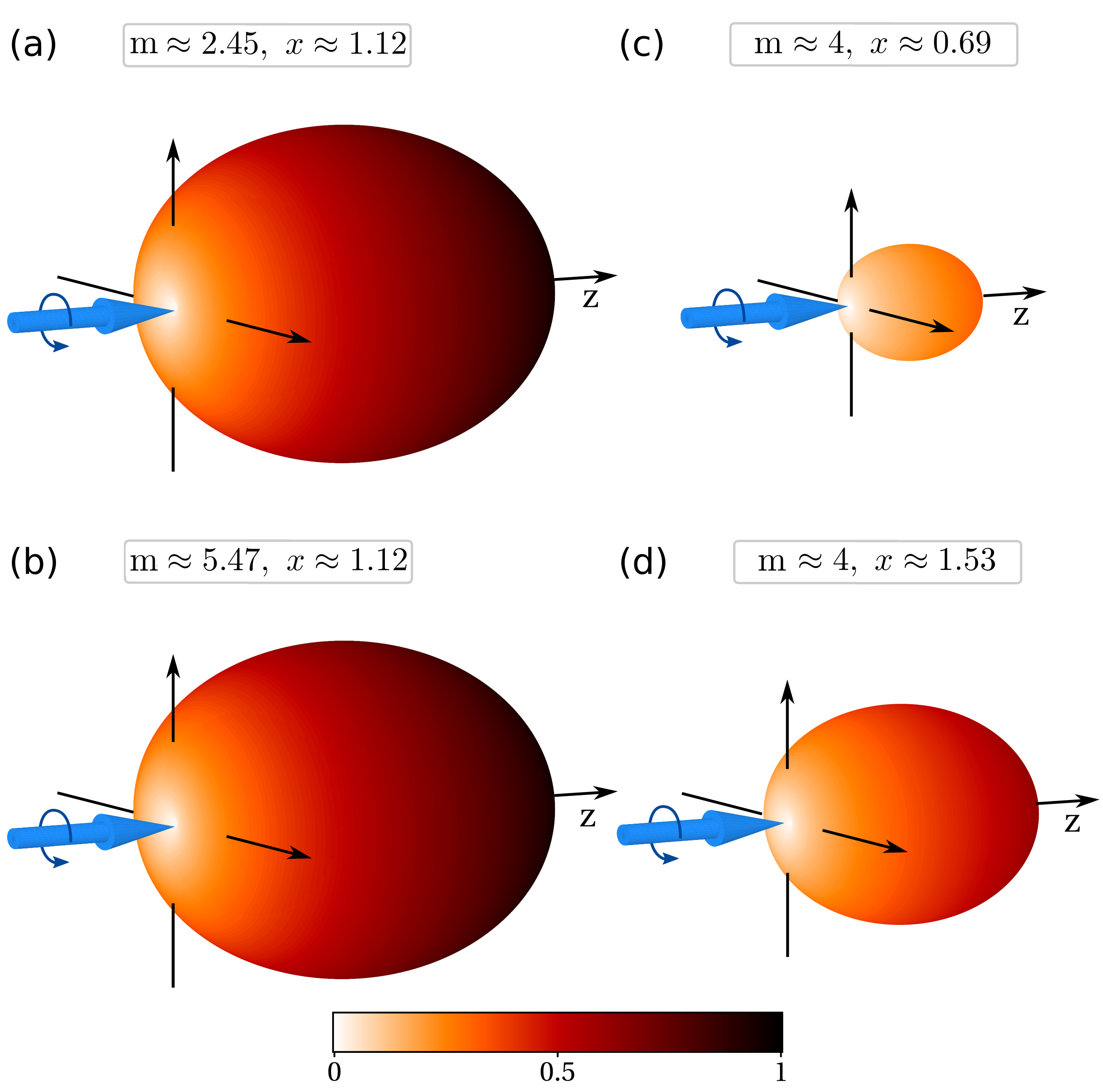}
\captionsetup{justification= raggedright}
\caption{Scattering radiation patterns arising from  dual materials in the dipolar spectral regime, normalized by the maximum value at the optimum forward light scattering condition: (a)-(b) two dual materials satisfying the optimum forward light scattering condition; (c)-(d) two Ge spheres with refractive index contrast ${\rm{m}} \approx 4$ below and above the quadrupole resonances, respectively.}
\label{radiation}
\end{figure}

Let us now briefly examine the so-called optimum forward light scattering condition, which was predicted in the limit of small particle for a particular diamond-like nano-sphere~\cite{aspnes1983dielectric} with refractive index contrast ${\rm{m}} \approx 2.45$~\cite{luk2015optimum,zhang2015dielectric}. We address this problem with a different approach that consists in maximizing the dipolar scattering efficiency (Eq.~\eqref{scattering} but retaining only $l=1$) at the first Kerker condition given by the dipolar node of second kind, i.e., $S'_1({\rm{m}},x)=0$, which yields:
\begin{align} \label{optimum}
\max \left[ {Q_{\rm{scat}}} \right] =12 \max{ \left|{{\frac{S'_1(x)}{x \xi'_1(x)}}} \right|}^2 \approx 3.75, && \text{with} && x_{\rm{max}} \approx 1.12.
\end{align}
Here $\xi_l(x) = x h_1(x)$, where $h_l(x)$ are the outgoing spherical hankel functions~\cite{bohren2008absorption}.
Equation~\eqref{optimum} shows that the dual scattering efficiency just depends  on the $x$ size parameter at the first Kerker condition and its maximum value is bounded. From the dipolar node of second kind evaluated at $x_{\rm{max}}$, namely, the size parameter that gives rise to the maximum dual scattering efficiency, it is straightforward to derive the material(s) that satisfy the optimum forward light scattering condition. Mathematically, this is equivalent to finding the set of refractive contrast indexes satisfying $S'_1({\rm{m}}x_{\rm{max}})=0$,
as shown graphically in Fig.~{\ref{lombriz}}.
Interestingly, for the asymptotic limit given by ${\rm{m}} \gg x, l,$ the  dipolar node of the second kind reads as $S'_1({\rm{m}}x) \approx \sin ({\rm{m}}x)$ and, then, the optimum forward light scattering condition is not a transcendental but an analytical solution given by the simple form ${\rm{m}}=q\pi / x_{\rm{max}}$, where $q \in \mathbb{R}$. 

For completeness, we illustrate the scattering efficiency radiation pattern, $dQ_{\rm{sca}} / d\Omega$ ~\cite{olmos2020optimal}, arising from two materials with ${\rm{m}}\approx 2.45$ and ${\rm{m}}\approx 5.47$ satisfying the optimum forward light scattering condition, in Fig.~\ref{radiation}a and Fig.~\ref{radiation}b, respectively. As previously discussed, the scattering radiation pattern is identical since the optimum forward light scattering condition arises at a fixed $x$ size parameter ($x_{\rm{max}} \approx 1.12$) for an infinite number of materials, contrary to interpretations~\cite{zhang2015dielectric,luk2015optimum}. Finally, let us discuss one aspect of interest about the dipolar regimes arising beyond the $m_{\rm{q}}$ and $e_{\rm{q}}$ resonances: the scattering radiation pattern arising from HRI dual spheres such as Ge considerably exceeds the one emerging from the same refractive index contrast in the limit of small particle, as can be inferred from Fig.~\ref{radiation}c and Fig.~\ref{radiation}d. This phenomenon could drive future experiments based in the EM duality restoration: EM helicity conservation and absence of backscattered light for HRI spheres in the dipolar spectral regime arising well-beyond the limit of small particle. 

In conclusion, we have unveiled a fundamental property of the scattering by a dielectric Mie sphere: the EM helicity can not be fully conserved in a multipolar scattering process. This finding precludes the ideal EM duality restoration and the perfect-zero optical backscattering condition in a multipolar scattering process.  The proof is general since it is solely based on a fundamental mathematical property of the Bessel functions and, thus, remains valid regardless of the particle size, refractive index contrast, incident wavelength, and multipole order. Nevertheless, we have shown that the almost entirely preservation of the EM helicity can be used as a probe of pure-multipolar spectral regimes, particularly, of dipolar nature beyond its presumed spectral region. This intriguing finding shows that the dipolar behaviour is not necessarily limited to small particles, showing that optical forces~\cite{chaumet2000time}, radiation pressure~\cite{nieto2010optical} and light transport effects~\cite{gomez2012negative}, originally derived from dielectric particles in the limit of small particle, can be extended for relatively large HRI spheres. Finally, we have proved that the optimum forward scattering condition, originally derived for a specific nano-particle, is satisfied for an infinite number of materials at a fixed $x$ size parameter. We firmly believe that the results of our calculations open new perspectives in the study of the scattering by dielectric Mie spheres, including new possible applications of HRI particles as building blocks in photonic devices.

The authors dedicate this work to the memory of their beloved colleague and friend, Prof. Juan José Sáenz, who passed away on March 22, 2020.

This research was supported by the Basque Government
(Project PI-2016-1-0041 and PhD Fellowship PRE-2018-
2-0252) and by the Spanish MINECO and MICINN and
European Regional Development Fund (ERDF) Projects:
FIS2015-69295-C3-3-P, FIS2017-
91413-EXP, FIS2017-82804-P, PGC2018-095777-B-C21 and
PhD Fellowship FPU15/ 03566.

\bibliography{New_era_18_06_2019}
\end{document}